\documentclass[fleqn,10pt]{wlscirep}
\usepackage[utf8]{inputenc}
\usepackage[T1]{fontenc}
\usepackage{graphicx}
\usepackage{xcolor}

\usepackage{pgf,pgfplots,tikz,subfig}
\usepackage{nicematrix}
  \usepackage{array,multirow,graphicx}
  \usepackage{csquotes}
  \usepackage{url}
  \usepackage{enumitem}
  \setlist[enumerate]{itemsep=0mm}
  \usepackage{hyperref}

\title{
Demystifying the COVID-19 vaccine discourse on Twitter
}

\author[a,*]{Zainab Zaidi}
\author[b]{Mengbin Ye}
\author[a]{Fergus John Samon}
\author[a]{Abdisalam Jama}
\author[a]{Binduja Gopalakrishnan}
\author[a]{Chenhao Gu}
\author[a]{Shanika Karunasekera}
\author[a]{Jamie Evans}
\author[a]{Yoshihisa Kashima}
\affil[a]{University of Melbourne, Melbourne, Australia}
\affil[b]{Curtin University, Perth, Australia}

\affil[*]{Corresponding author:Zainab.RaziaZaidi@unimelb.edu.au} 

\keywords{COVID-19 vaccine, vaccine hesitancy, anti-vax, stance detection, topic modelling, misinformation \vspace{-2mm}}

\begin{abstract}
Developing an understanding of the public discourse on COVID-19 vaccination on social media is important not only for addressing the current COVID-19 pandemic, but also for future pathogen outbreaks. We examine a Twitter dataset containing 75 million English tweets discussing COVID-19 vaccination from March 2020 to March 2021. We train a stance detection algorithm using natural language processing (NLP) techniques to classify tweets as `anti-vax' or `pro-vax', and examine the main topics of discourse using topic modelling techniques. While pro-vax  tweets ($37$ million) far outnumbered anti-vax tweets ($10$ million), a majority of tweets from both stances (63\% anti-vax and 53\% pro-vax tweets) came from \textit{dual-stance} users who posted both pro- and anti-vax tweets during the observation period. Pro-vax tweets focused mostly on vaccine development, while anti-vax tweets covered a wide range of topics, some of which included genuine concerns, though there was a large dose of falsehoods. %A number of topics were common to both stances, though pro- and anti-vax tweets discussed them from opposite viewpoints. 
Memes and jokes were amongst the most retweeted messages. Whereas concerns about polarisation and online prevalence of anti-vax discourse are unfounded, targeted countering of falsehoods is important. %Future research should examine the role of memes and humours in driving online social media activities.
\vspace{-2mm}
\end{abstract}
\begin{document}

\flushbottom
\maketitle
\thispagestyle{empty}

\section*{Introduction}
%\dropcap{D}
Discourse on vaccination began early during the COVID-19 pandemic and has since received sustained attention in mainstream media, social media, and among the general public around the world. Although the widespread uptake of COVID-19 vaccines in many countries has enabled them to shift towards `living with COVID', the emergence of new variants and the inevitable waning of immunity mean that vaccination remains central to the world's capability of coping with future infection waves \cite{Murray2022}. At this juncture, understanding the public vaccination discourse and its dynamics is critically important for governments, policymakers, and scientists to maintain and increase trust and uptake of vaccines in the future. By conducting a comprehensive study of English-language Twitter activities, we attempt to demystify the online vaccination discourse and provide a better understanding of its content and dynamics.

Existing research on COVID-19 vaccine discourse has focused on anti-vaccination misinformation, examining its spread over social media, and its influence on vaccination debate \cite{Xu2021,Fazel2021,Gori2021,Hayawi2022,Yousefinaghani2021,Muric2021}. Mainstream media have frequently reported on, and fact-checked, misinformation and falsehoods related to COVID-19 vaccines \cite{ABC-aborted,BBC-dna,NPR-moralvaccine,NYTimes-cellline,Plandemic-nytimes,Reuters-dna,Reuters-fertility,Reuters-nano}. In the course of the two years, a perception has emerged that discussion around COVID-19, and especially about vaccination, is highly polarised. In this view, anti- and pro-vaccination discourses run in parallel without interacting with each other, each coalescing around shared narratives while ignoring information and arguments that challenge them~\cite{Schmidt2018,Cinelli2020}. 

Contrary to this popular perception, we are yet to have a clear picture of the content and dynamics of the public discourse on COVID-19 vaccines on social media largely due to the absence of a reliable and scalable method for stance detection, namely, measuring pro- or anti-stance on an issue expressed in social media messages. To begin, we do not know whether pro- or anti-vax messages predominate the online discourse. On the one hand, manual coding \cite{Gori2021,Hou2021} has found relatively more anti- than pro-vax tweets, but this method is unscalable and limited to a few thousand randomly sampled tweets. On the other hand, scalable unsupervised learning methods, such as sentiment analysis and topic modelling, are not always reliable and can yield contradictory results. Whereas studies using sentiment analysis have reported prevalent positive sentiment about vaccines over Twitter \cite{Yin2014,Lanyi2022,Fazel2021,Yousefinaghani2021}, a topic modelling approach to stance prediction in \cite{Yousefinaghani2021} classified relatively more tweets as anti-vax than pro-vax. 

Similarly, it is still unclear whether vaccination discourse is indeed polarised. Although human messaging, rather than bot activities, appears to shape the online discourse \cite{Xu2021}, the most prolific of those human contributors do not appear to show a strongly polarised vaccination stance. Gori et al. \cite{Gori2021} found this to be the case based on manually annotated Italian tweets around the time of their vaccination campaigns. Intriguingly, however, these users' tweet contents were extremely polarised. The trend of sentiments further complicates the picture, with \cite{Greyling2022} reporting a global downward trend in positive sentiments towards vaccination over 6 months. Does this mean anti-vaccination discourse is gaining the upper hand? 

Further, if indeed the online discourse is responsible for changing sentiments, we still do not know what (mis)information and arguments are affecting vaccination stances. To be sure, past studies have suggested that safety, mistrust of government and pharmaceutical companies, accessibility issues, conspiracies and misinformation are key barriers to vaccine uptake \cite{Lanyi2022,Kucukali2022,Hou2021}. Topic modelling of vaccine-related tweets \cite{Lyu2021} identified opinions about vaccination, vaccine progress, and instructions on getting vaccine as main topics. However, neither study examined vaccination stance, and therefore could not examine which topic was associated with which stance. To the best of our knowledge, only one study \cite{Hayawi2022} used a BERT (Bidirectional Encoder Representations from Transformers) model for stance detection of COVID-19 vaccine related tweets, trained with 15,000 labelled tweets. They showed the efficacy of the model in stance detection; however, they did not examine the topics being discussed.

To gain a comprehensive understanding of the content and dynamics of social media discourse on COVID-19 vaccination, we examine $75$ million English language tweets related to COVID-19 vaccine. These tweets are extracted, using vaccine related keywords over a yearlong observational period from Mar.\ 20, 2020 to Mar.\ 23, 2021, from a publicly available dataset of COVID-19 tweets collected by R. Lamsal~\cite{Lamsal2021}. We use Natural Language Process (NLP) and OpenAI's GPT transformer-based stance detection tool \cite{Rao2019,Radford2018}, trained on a subset of $42,000+$ manually labelled tweets, to classify anti-vax and pro-vax tweets. Both classes are then analysed to determine discussion topics using a composite strategy involving the GS-DMM topic modelling tool~\cite{Yin2014} and a manual search to compile a comprehensive list of topics and relevant keywords and phrases. The details of the method are in the \hyperref[methods]{Materials and Methods}. 

%\vspace{-3mm}
\section*{Results}

\begin{figure*}[tbh!] 
    \subfloat[]{
        \includegraphics[trim={0 55 0 0},width= 0.5\textwidth,height=5.5cm]{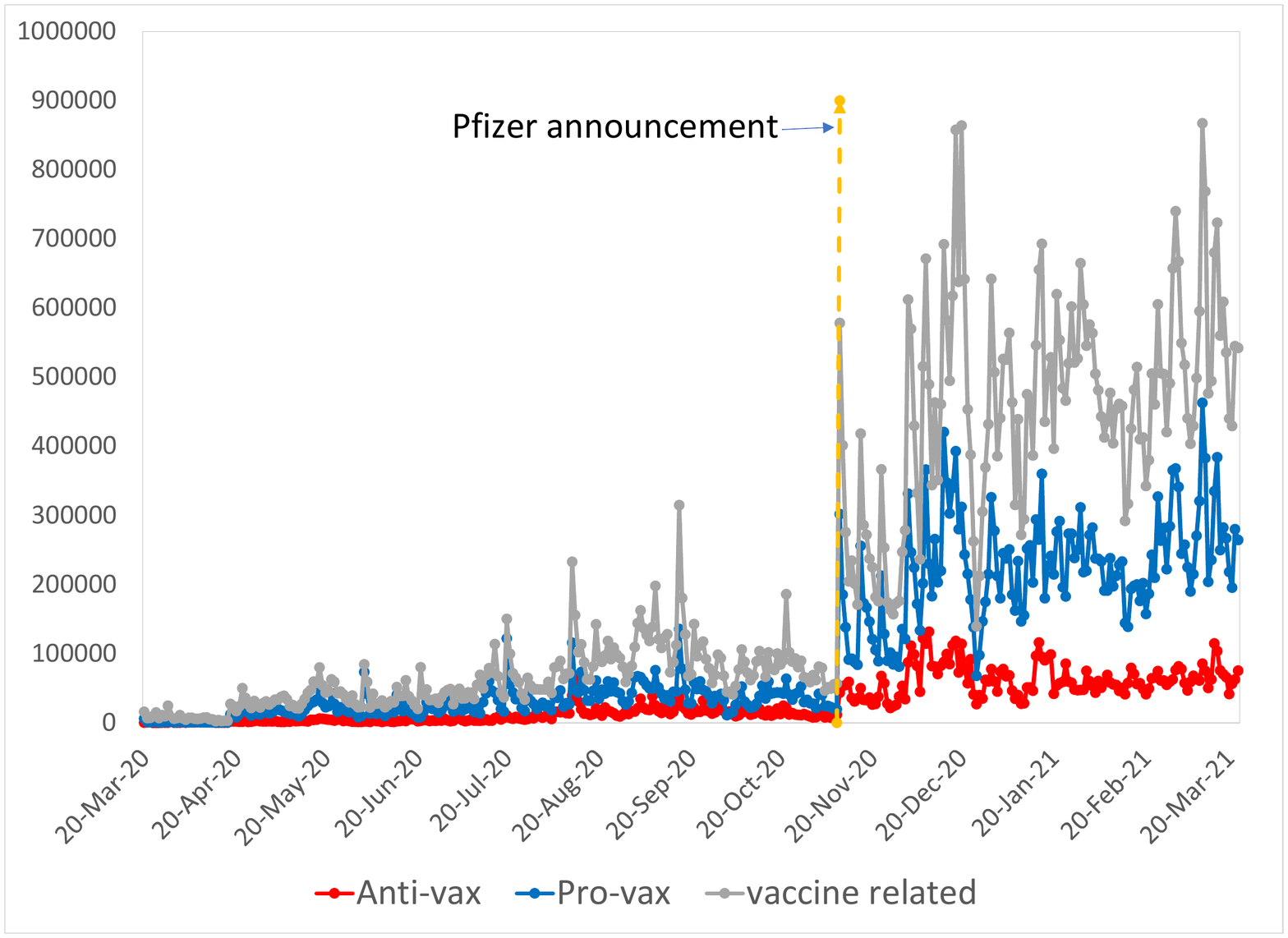}
        \label{result_Lamsal}}
        \hfill
    \subfloat[]{
        \includegraphics[trim={100 55 100 55},width=0.5\textwidth,height=5.5cm]{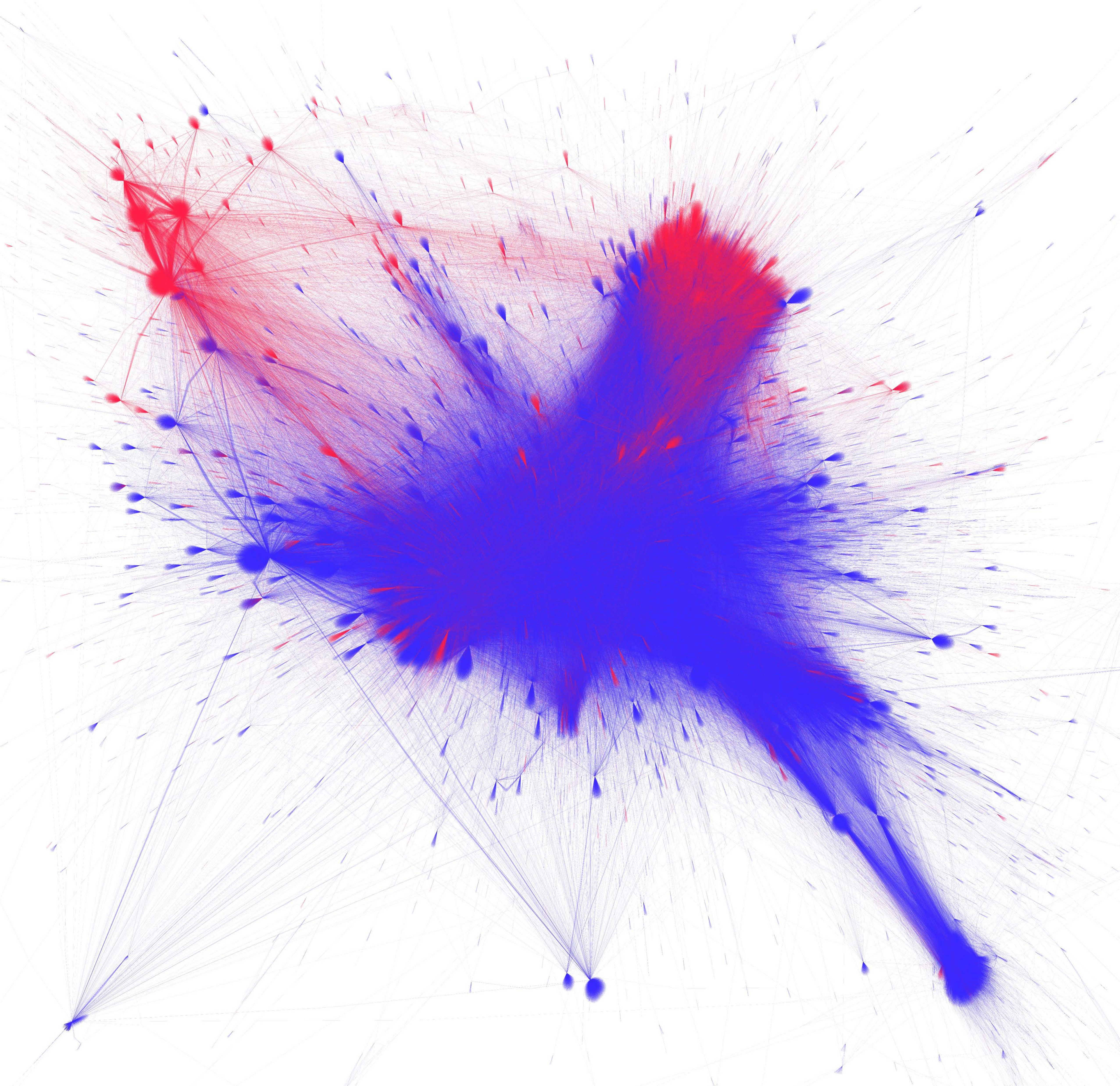}
        \label{retweet-network}}\vspace{-3mm}
        \hfill \\
    \subfloat[]{
        \includegraphics[width= 0.32\textwidth,height=4.8cm,scale = 0.6]{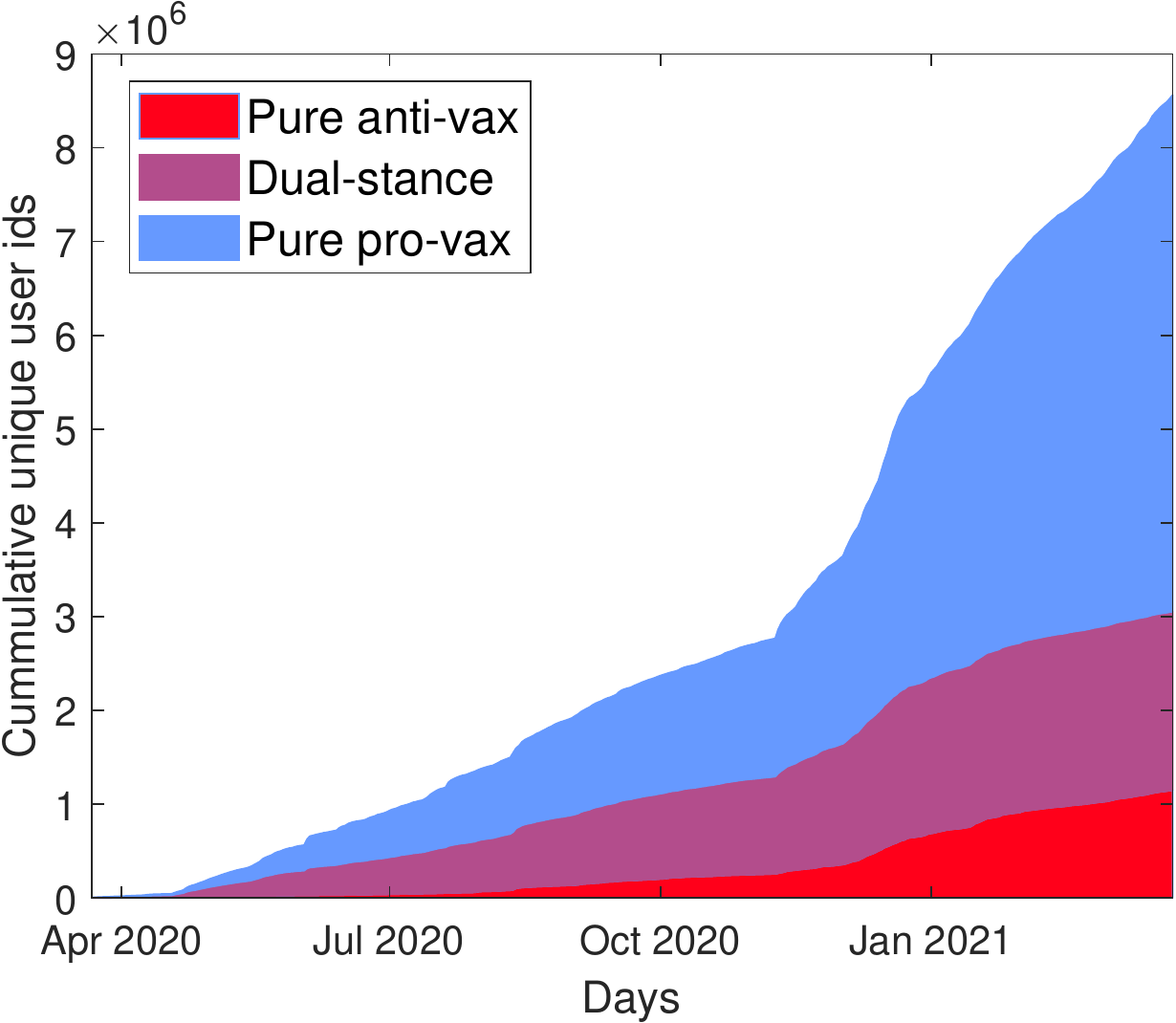}
        \label{user_growth}}
    \hfill
    \subfloat[]{
        \includegraphics[width= 0.32\textwidth,height=4.7cm,scale=0.6]{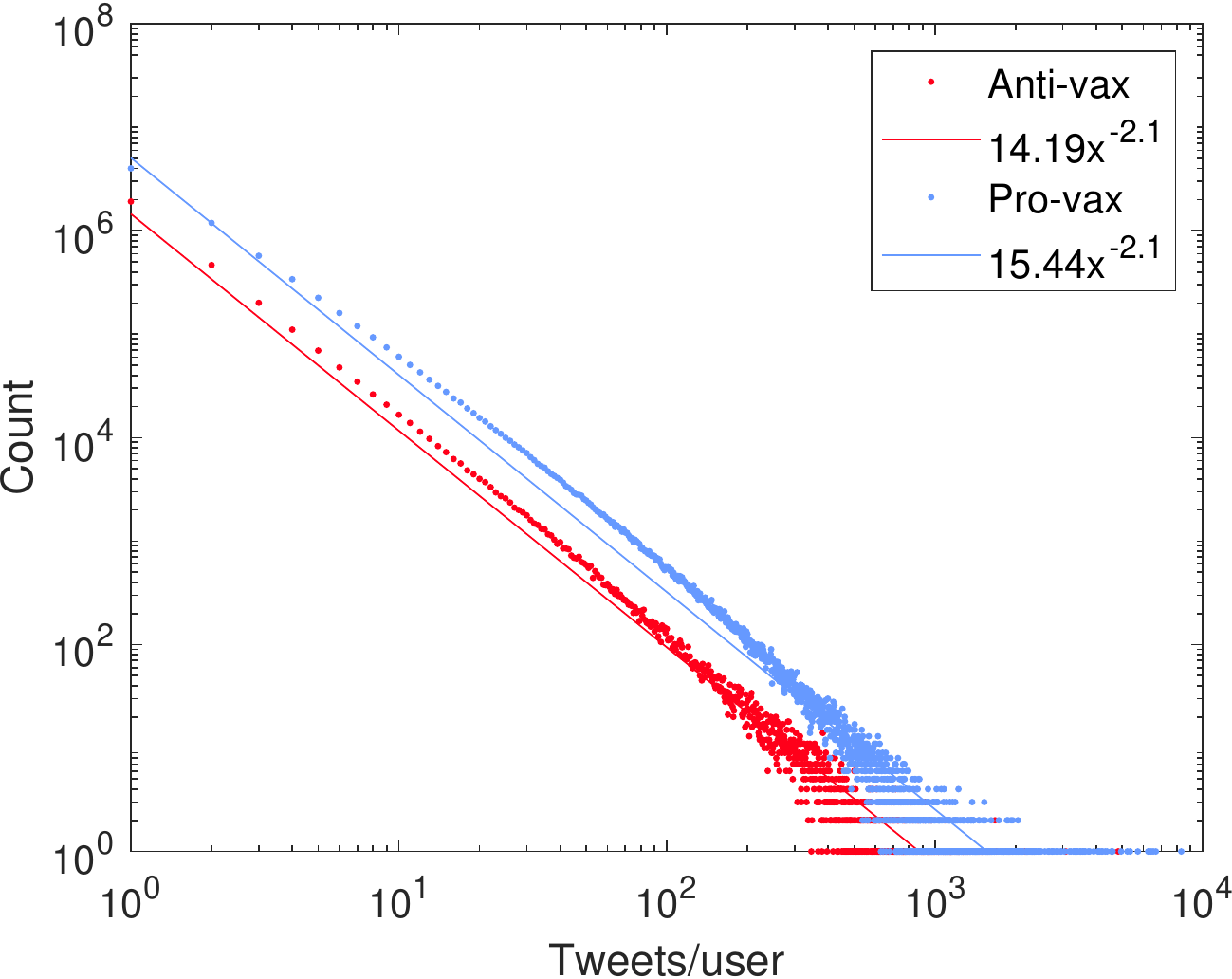}
        \label{power-law}}
        \hfill
    \subfloat[]{
        \includegraphics[width= 0.32\textwidth,height=4.7cm,scale=0.6]{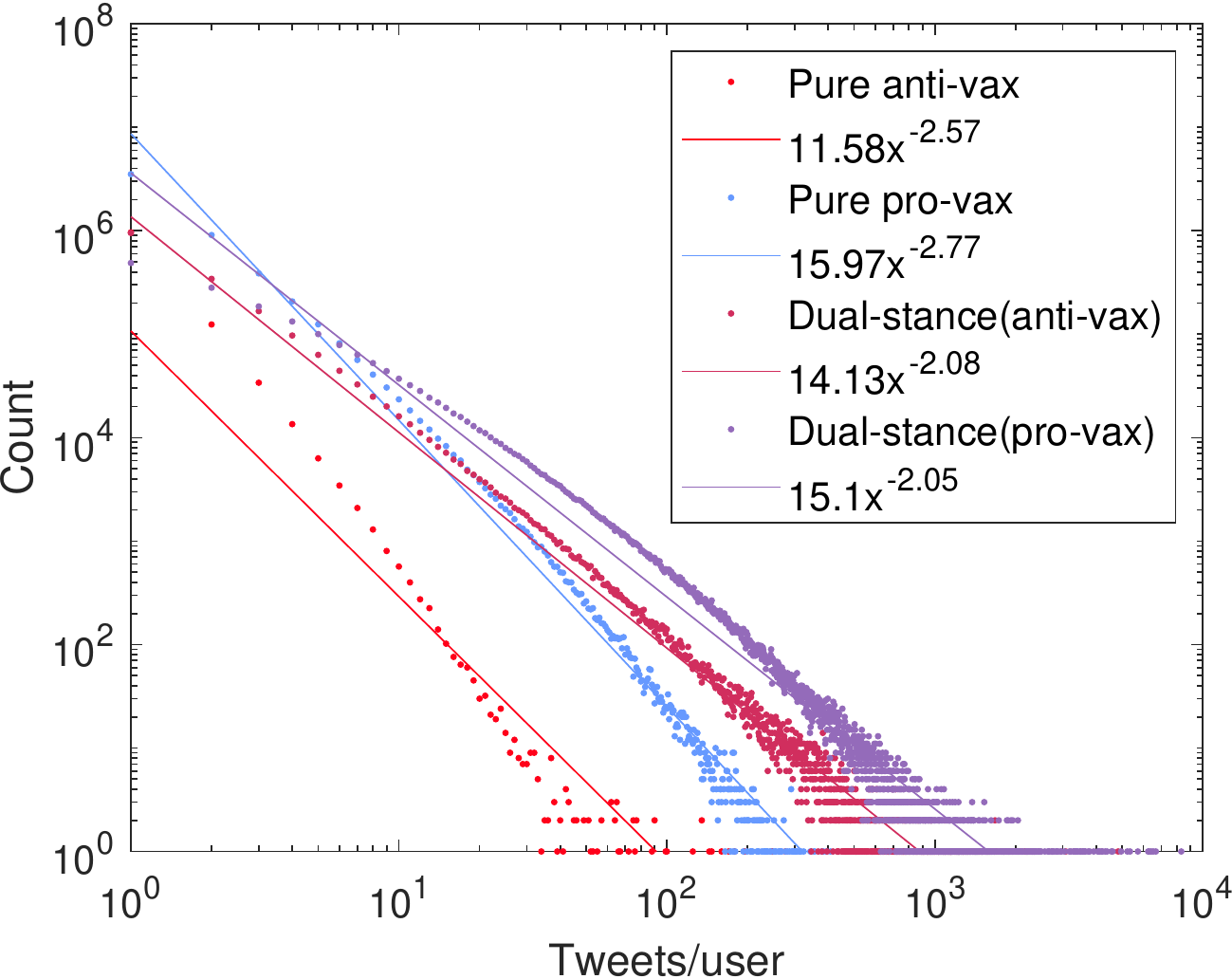}
        \label{power-law-all}}\vspace{-3mm}
    \caption{One year of COVID-19 vaccine tweets, from Mar.~20, 2021 till Mar. 23, 2021, are analysed in this study. The tweets are classified into anti-vax and pro-vax tweets through NLP-based stance detection. 
    A 10-fold increase in vaccine-related tweets is observed after Nov. 9, 2020, when Pfizer announced the results of their preliminary analysis \cite{Pfizer-announcement}. 
    (a) Time-series of total COVID-19 vaccine related tweets from Mar.\ 2020 to Mar.\ 2021 and its breakdown into anti-vax and pro-vax tweets. (b) The network of users on Twitter, with links connecting the retweeting users to the users who posted the original tweet. Blue (red) link represent a pro-vax (anti-vax) retweet, while violet links represent both pro-vax and anti-vax tweets originated by one and retweeted by the other user. Retweets classified as `neutral/irrelevant' are not included.  (c) The number of unique user IDs throughout the year for pure anti-vax, dual-stance, and pure pro-vax groups steadily increased, with a higher rate of increase in the post-vaccine period. (d) The count of users with a given number of tweets, over a log scale. Tweets/user followed a power law $\alpha x^\gamma$ distribution for both classes (also shown with straight lines), i.e., there were many users with few tweets and there were a small number of users with many tweets in both groups. (e) Tweet/user and associated power law distribution drawn separately for dual-stance users for anti-vax and pro-vax tweets and pure anti-vax and pure pro-vax users. Dual-stance users had more users with many tweets than the pure anti-vax and pure pro-vax groups.}
    \vspace{-5mm}
\end{figure*}

Our stance detection algorithm shows that pro-vax tweets clearly outnumbered anti-vax tweets throughout the observation period (Fig.~\ref{result_Lamsal}): $37,047,378$ pro-vax, $10,567,955$ anti-vax, and $28,322,526$ neutral or irrelevant tweets. Moreover, of the total of $8,563,466$ unique user IDs (for anti-vax and pro-vax tweets), $5,509,840$ ($64.3\%$) tweeted only pro-vax messages, but $1,151,726$ ($13.4\%$) tweeted only anti-vax messages. Most intriguingly, $1,901,900$ ($22.2\%$) {\it dual-stance} users sent out both pro- and anti-vax tweets (Fig.\ \ref{user_growth}), whose presence was hinted at by \cite{Gori2021}. The communication network, shown in Fig.~\ref{retweet-network}, where links represent retweets, while nodes represent users (see \hyperref[methods]{Materials and Methods}), shows some like-minded clusters, but they are far from isolated. If anything, it only shows a small anti-vax cluster.

Two observations stand out. First, pro-vax discourse is predominant in the English-language Twittersphere, with more than $80\%$ of the users
posting a pro-vax tweet. Second, we find limited evidence of polarised discourse. A majority of those who sent an anti-vax tweet also sent out a pro-vax tweet ($62.2\%$). Rather, there are sizable dual-stance users who engage with both pro- and anti-vax contents. In fact, they are the most prolific contributors, who sent the majority of the pro- and anti-vax tweets. Dual-stance users dominate pro-vax tweets. They constitute only a quarter of the total pro-vax users, but contributed 53\% of the pro-vax tweets. In contrast, dual-stance users' contributions of anti-vax tweets are proportionate to those by their pure anti-vax counterparts. They are 62\% of anti-vax users and contributed 63\% of the anti-vax tweets (see details in Supplementary Information). 

Using topic modelling, we identify that, although a single topic -- `vaccine development' -- dominates pro-vax tweets, anti-vax tweets mention more diverse topics, including `vaccine mandates', `vaccine side effects', `masks and lockdowns'. Some topics commonly appeared in both anti- and pro-vax tweets (e.g., `masks and lockdowns', `Pfizer/Moderna vaccines') with opposing perspectives. Memes and humorous tweets (classified as the topics `jokes' and `jokes (side effects)') were among the most retweeted messages for both the anti- and pro-vax tweet sets, with 7 of the 10 most retweeted anti-vax tweets involving memes or jokes. In fact, they dominated the tweets of those who tweeted only pro- or anti-vax messages. The majority of the anti-vax tweets contained falsehoods (e.g., `COVID-19 is a hoax', `COVID-19 is a secret plan to limit people's rights') and only $10-15\%$ expressed genuine concerns (e.g., mandatory vaccination, blood clots following the AstraZeneca (AZ) vaccine).

%\vspace{-3mm}
\subsection*{Overview of the Vaccine Debate}

Figure \ref{result_Lamsal} presents the yearlong time series of the volume of tweets classified as pro-vax (blue) and anti-vax (red) and the total number of tweets (grey). The first notable feature is a $10$-fold increase in vaccine related tweets from Nov.\ 9, 2020, which coincides with Pfizer announcing preliminary efficacy results for their COVID-19 vaccine~\cite{Pfizer-announcement}. We use this announcement date to divide the year into two significantly different periods, viz. the `{\it pre-vaccine}' and `{\it post-vaccine}' periods. The second significant aspect is that the number of pro-vax tweets is over 3 times than anti-vax tweets, noting that the pro-vax tweets contain tweets from governments, pharmaceutical companies, and media agencies, besides regular users. 

Figure~\ref{user_growth} shows the growth of unique user IDs associated with pure anti-vax (red), dual-stance (violet), and pure pro-vax (blue) tweets over the year, while Fig.~\ref{power-law} shows the number of users with a given number of tweets during the study period. Both anti-vax and pro-vax classes follow power law distribution with similar exponent. This observation conforms to the previous studies \cite{Lu2014,Bild2015}, which found user aggregate behaviour over Twitter often follows power law distributions. Figure \ref{power-law-all} plots separately the tweets/user statistics for dual-stance users from pure anti-vax and pure pro-vax users, which also shows that dual-stance users are more active than those who posted only single stance tweets. Moreover, both classes of anti-vax and pro-vax also exhibit almost identical user growth rates -- lower and higher in the pre-vaccine and post-vaccine periods, respectively. However, pro-vax tweets have a higher growth rate than anti-vax tweets. 

\begin{figure*}[tbh!]
\subfloat[]{
        \includegraphics[trim={0 8 0 0},width= 0.5\textwidth,height=10cm]{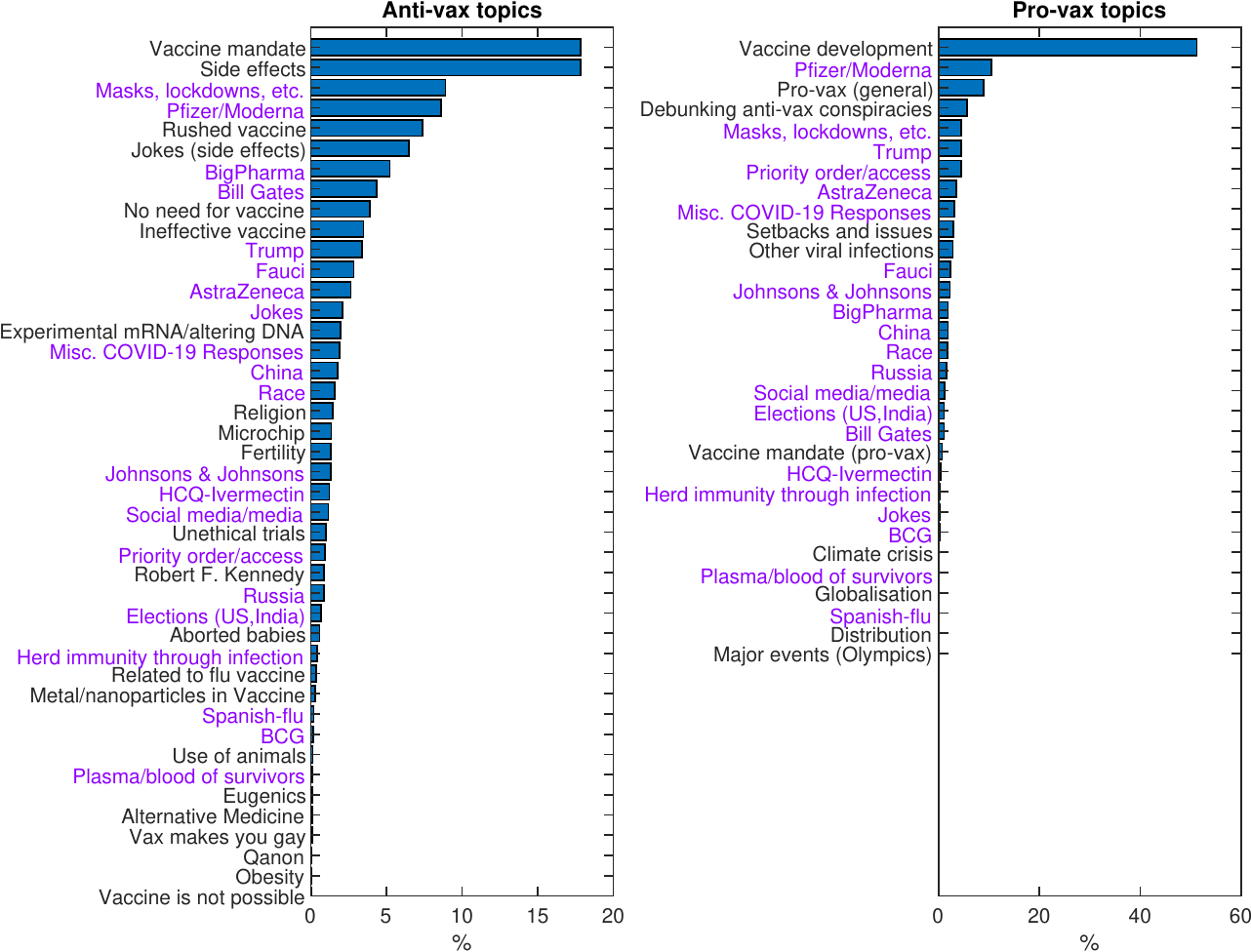}
        \label{topics_figure}}
        \hfill
        \subfloat[]{
        \includegraphics[trim={0 8 0 0},width= 0.5\textwidth,height=10cm]{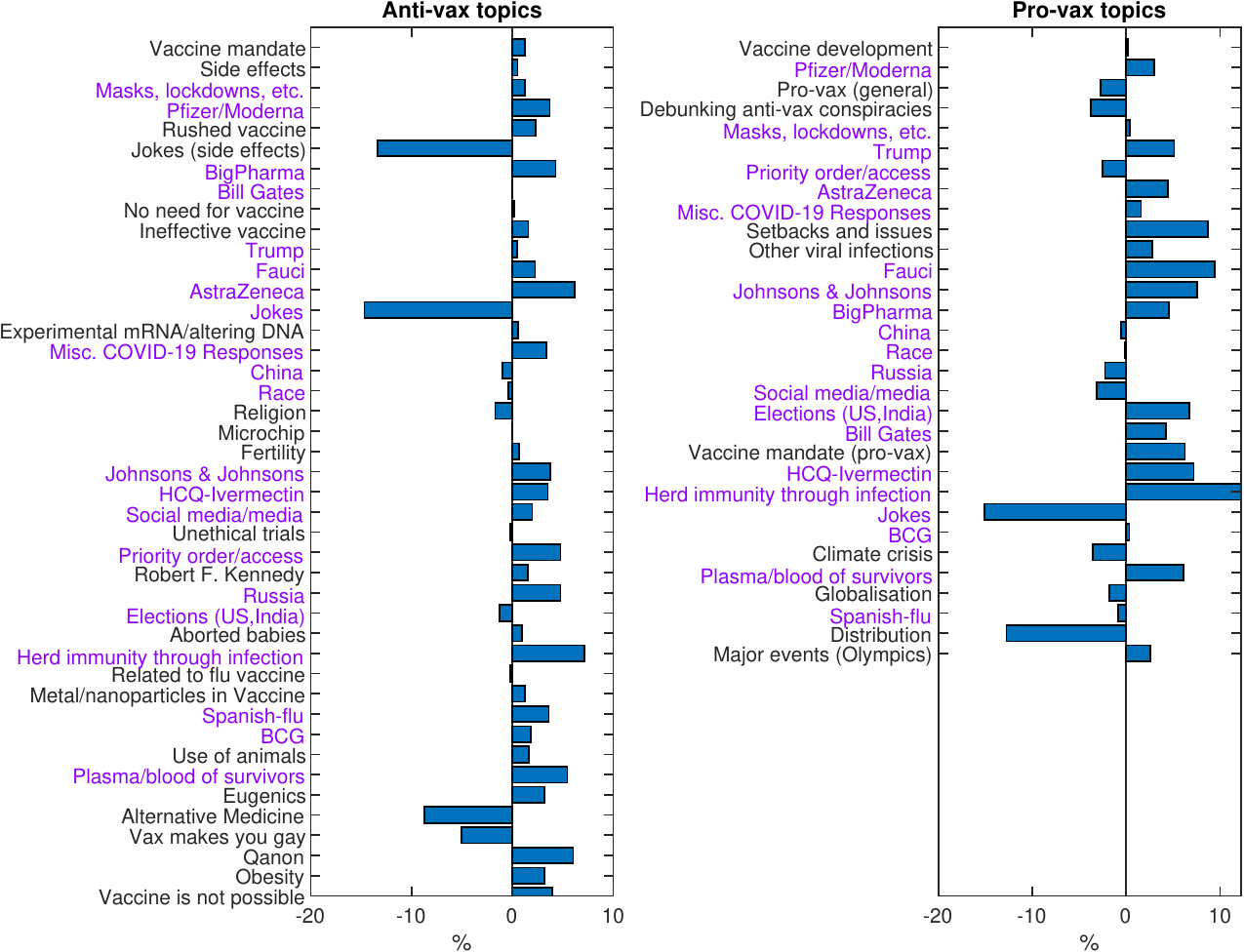}
        \label{dual_dev}}
        \vspace{-3mm}
\caption{Using topic modelling techniques and manual search, 43 topics are found for anti-vax tweets and 31 topics are found for pro-vax tweets, with 21 common topics (shown in violet colour). a) The topics that are identified in the anti-vax and pro-vax tweets, ranked according to the fraction of tweets containing the given topic. b) The normalised difference in actual versus expected (according to the population ratio) contributions of dual-stance users for each topic. A positive (negative) value indicates that tweets from dual-stance users contributed more than expected (less than expected) in relation to contributions from users who only tweeted anti-vax/pro-vax. Dual-stance users are relatively more active in the majority of the discussion topics. Interestingly, jokes are more popular among pure anti-vax and pure pro-vax users.}
\vspace{-6mm}
\end{figure*}

Using topic modelling, we identify 53 distinct discussion topics, with 31 and 43 topics exclusive to pro- and anti-vax tweets, respectively. 21 topics are common across both classes. Note that the topics are not mutually exclusive, and one tweet can contain multiple topics, while $20-25\%$ of both pro- and anti-vax tweets were not classified into any topic. Figure~\ref{topics_figure} shows the percentages of tweets for the identified topics, relative to the total number of pro- and anti-vax tweets. One striking difference is the diversity of topics covered by anti-vax tweets. Whereas `vaccine development' ($51.3\%$) dominates pro-vax tweets,  the anti-vax tweets cannot be summarised with a few topics. The top anti-vax theme ($29.5\%$) is 'vaccine safety', a combination of `side effects', `rushed vaccine', and `jokes (side effects)', followed by `vaccine mandate' ($17.9\%$), and all other topics with a less than $10\%$ share. Note that we separate the humorous tweets into two groups, with jokes regarding side effects in `jokes (side effects)' and the rest in `jokes', mainly to quantify the safety concerns.% in terms of tweet's volume.

To characterise dual-stance users' tweets relative to pure anti- and pure pro-vax users, we compute the normalised difference between the actual and the expected contributions of dual-stance users in each topic, as shown in Fig.\ \ref{dual_dev}. The expected contribution of dual-stance users for a pro-vax (anti-vax) topic is calculated by multiplying the number of topic tweets by the ratio of dual-stance pro-vax (anti-vax) tweets to total pro-vax (anti-vax) tweets, i.e., $0.53$ ($0.63$). An excess contribution by dual-stance users corresponds to lesser contribution, by equal amount, from the pure pro-vax/anti-vax users. The most striking observations are: 1) dual-stance users are more active than both pure pro-vax and anti-vax users in the majority of the discussion topics, and 2) jokes are mostly popular with both the pure pro- and anti-vax users (see Supplementary Information for more detail). 

Next, we examine the key topics\footnote{Details about less significant topics are in Supplementary Information.} and their dynamics over time by plotting daily tweets and linking their significant peaks to COVID-19 related current events. 

%\vspace{-2mm}
\subsection*{The Anti-vax Tweets}\label{sec:anti}

\begin{figure*}[tbh!]
    \subfloat[Anti-vax topics]{
        \includegraphics[width= 0.46\textwidth,height=7cm]{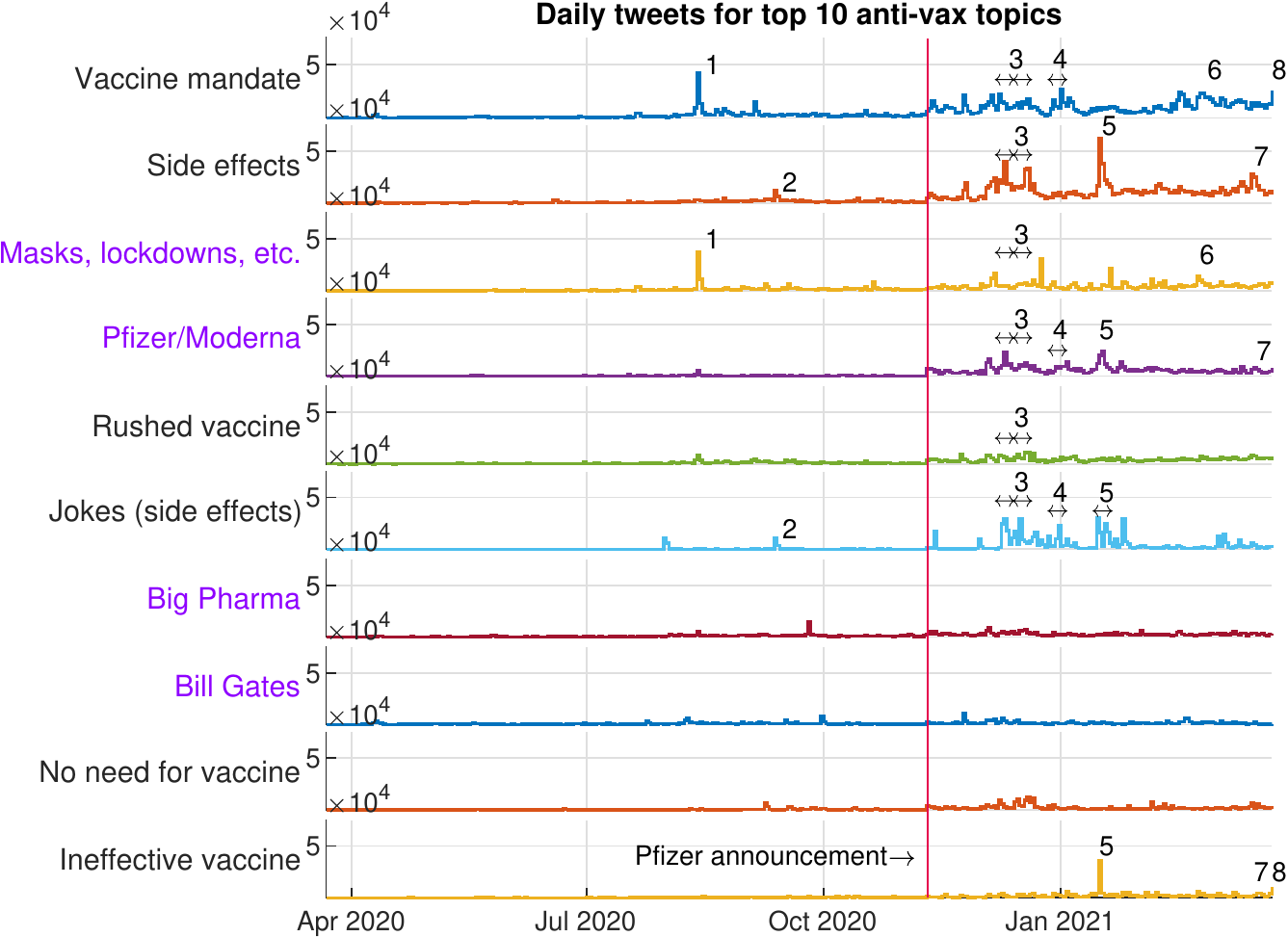}
        \label{anti-topics}}
    \hfill
    \subfloat[Pro-vax topics]{
        \includegraphics[width= 0.52\textwidth,height=7cm]{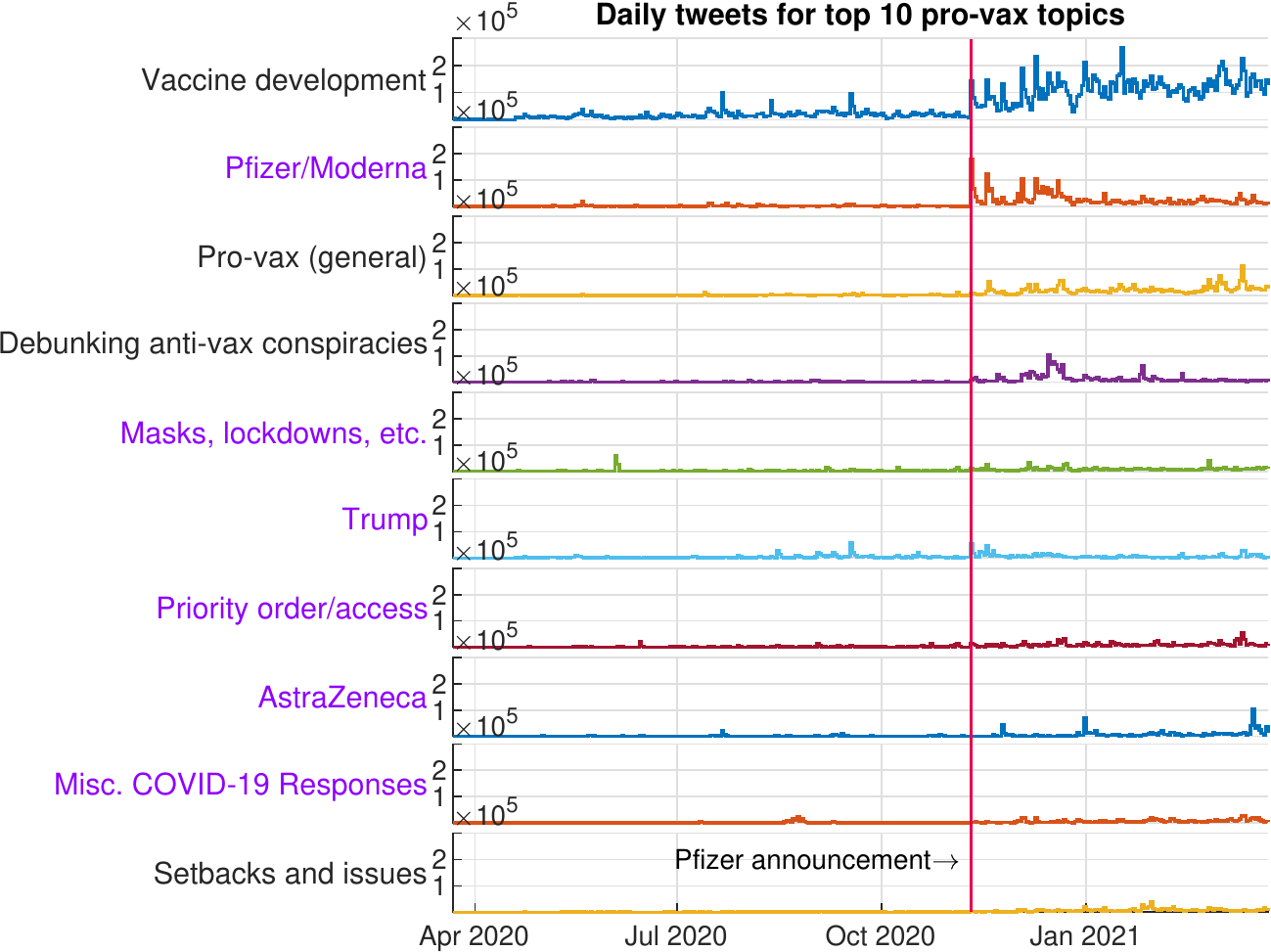}
        \label{pro-topics}}\vspace{-3mm}
    \caption{`Vaccine mandate' and the broad theme of safety, which includes `side effects', `rushed vaccine', `jokes (side effects)', are the most significant topics in anti-vax discussions, whereas the pro-vax tweets are dominated by updates around `vaccine development'. November 9, 2020 marks the start of the post-vaccine period with Pfizer announcing the preliminary results of their trials. (a) The top 10 topics in anti-vax class and (b) the top 10 topics in pro-vax class.}
    \vspace{-5mm}
\end{figure*}

We start with anti-vax tweets. Despite their relative infrequency, anti-vax tweets appear to drive many significant dynamics of vaccine discourse. The peaks numbered in Fig.~\ref{anti-topics} of the top 10 anti-vax topics appear to be linked to COVID-19 related current events.\vspace{-1mm} 

\begin{enumerate}
    \setlength\itemsep{-0.5em}
    \item {\bf Aug.\ 12, 2020} A peak in the topic `vaccine mandate' occurred one day after clinical outcomes for the Russian vaccine Sputnik were announced.
    \item {\bf Sep.\ 11, 2020} AstraZeneca's phase 3 trials were put on hold on Sep.\ 8, 2020, after one volunteer developed an unknown reaction~\cite{AZ-phase3}. Peaks are observed in the topics of `side effects' and `jokes (side effects)'.% on Sep.\ 11, 2020.
    \item {\bf Dec.\ 7-20, 2020} 
    Multiple peaks are observed in the topics of `side effects', `Pfizer/Moderna', `jokes (side effects)', `rushed vaccine', `vaccine mandate', and `masks/lockdowns', throughout the two weeks of Dec.\ 2020, coinciding with the Pfizer-BioNTech and Moderna vaccines gaining various approvals in the United Kingdom (UK) and from the USA's Food and Drug Administration (FDA)~\cite{Pfizer-approved-UK,Pfizer-eua,Pfizer-recommend-FDA,Moderna-recommend-FDA,Moderna-eua}, and the launch of UK's public vaccination program on Dec.\ 8, 2020. 
    %The news of a nurse fainting after receiving Pfizer vaccine in Tennessee ad a significant role in the peak on Dec.\ 18, 2020\footnote{\url{https://www.local3news.com/local-news/whats-trending/first-doses-of-covid-19-vaccines-administered-at-chattanooga-hospital-on-thursday/article\_922c71df-8abe-5c18-b6d0-8dafe716e8be.html}}. 
    \item {\bf Dec.\ 28, 2020-Jan.\ 3, 2021} Several countries closed their borders to the UK in response to the Alpha variant of COVID-19 on Dec.\ 21, 2020~\cite{UK-variant}, while the UK government changed the gap between Pfizer doses and approved the AZ vaccine on Dec.\ 30, 2020~\cite{UK-change-gap,AZ-approved}. We observe higher tweet activity for `vaccine mandate', `Pfizer/Moderna', and `jokes (side effects)' on multiple days including Dec.\ 31, 2020. 
    \item {\bf Jan.\ 15-17, 2021} A viral meme, about zombie apocalypse as a result of failed vaccine in the movie `I am Legend' and its relevance to the current pandemic, drove peaks in the `side effects', `jokes (side effects)', `Pfizer/Moderna', and `ineffective vaccine' topics on Jan.\ 15, 2021. On Jan.\ 16-17, 2021, the most significant news shared under the `jokes (side effects)' and `Pfizer/Moderna' topics was about 23 deaths in Norway post vaccination~\cite{Norway-deaths}. This incident is discussed in detail in the sequel.
    \item {\bf Feb.\ 22-26, 2021} UK's review of vaccine passports on Feb.\ 23, 2020~\cite{UK-VacPassport}, may be behind the spikes in `vaccine mandate' on Feb.\ 24 and 26, 2021. On Feb.\ 21, 2021, Dr.~Fauci's stated that Americans may need to wear masks in 2022~\cite{Fauci-masks}, which might be linked to a peak in the plot of `masks, lockdowns, etc.' the next day. 
    \item {\bf Mar.\ 11-17, 2021} On Mar.\ 11, 2021, use of the AZ vaccine was suspended across many countries due to fears of blood clots~\cite{AZ-banned}, causing peaks in `side effects' on Mar.\ 15, 2021 and `Pfizer/Moderna' on Mar.\ 16, 2021. Coincidentally, the third COVID-19 wave was sweeping across Europe~\cite{Third-wave} and there were multiple spikes in the `ineffective vaccine' and `vaccine mandate' topics.
\end{enumerate} 
%\vspace{-2mm}

Whereas it is challenging to establish a causal link between a possible triggering current event and the topics discussed in the tweet dataset, many of the anti-vax tweets appear to be spontaneous responses to key COVID-19 vaccine events. This suggests that it may be challenging to develop counter-narratives in a timely manner and deploy them to {\it prebunk} vaccine-related misinformation~\cite{ecker2022psychological}. Moreover, memes and funny tweets seem to play a significant role in progressing anti-vax narratives, and we will return to this in detail later. Next, we discuss the major anti-vax topics in detail.

%\vspace{-2mm}
\subsubsection*{Safety (Side Effects, Rushed Vaccine, Jokes (Side Effects))}
\label{safety}

Under the theme of `safety', we group three topics, `side effects', `rushed vaccine', and `jokes (side effects)', taking up $29.5\%$ of the anti-vax tweets. Of these, `side effects' accounts for the largest number of tweets, %, with a sustained presence after week 15 that in fact grew substantially in the post-vaccine period. Notably, `side effects' had significant peaks throughout the yearlong observation period.
discussing a broad range of illnesses due to receiving a vaccine dose, ranging from allergic reactions to death. Although some conspiracies found their way into the conversation, such as vaccines causing infertility, the tweets included some elements of true adverse reactions from vaccine trials. Nonetheless, many were exaggerations of facts, quoting facts out-of-context, and even falsehoods regarding incidents which never occurred (see the Supplementary Information for details). For example, media coverage on the 23 elderly people who died in Norway after taking the Pfizer vaccine in Jan.\ 2021~\cite{Norway-deaths,Norway-deaths2} was shared in 12,249 tweets and caused a peak in `Pfizer/Moderna' and `jokes (side effects)' topics in Fig.~\ref{anti-topics} on Jan.\ 16-17, 2021. An investigation by Norwegian Medicines Agency, reported by~\cite{Norway-deaths}, stated `common adverse reactions to mRNA vaccines may have contributed to worsening of their underlying diseases and a fatal outcome in some frail patients' -- this article was subsequently shared by 3,323 tweets that suggested vaccines `may have led' to deaths. Shortly after, our dataset showed thousands of tweets discussing an increase in the number of deaths in Norway.

The plot of `rushed vaccine' saw consistent presence in the post vaccine period. The peak on Aug.\ 12, 2022 is possibly linked to the announcement of the Sputnik vaccine. The tweets were typically either falsehoods or cherry-picked information taken out of context, highlighting the extremely rapid development and approval processes in contrast to previous vaccine development efforts. These doubts %expressed over the fast-tracking measures 
can relate to the mistrust over authorities and pharmaceutical companies reported by \cite{Lanyi2022,Kucukali2022}. An example is the discussion about Hydroxychloroquine (HCQ) (see Supplementary Information for details), which was banned as a possible COVID-19 treatment after multiple scientific studies \cite{HCQ}, but the anti-vax tweets suggested that HCQ was banned only to give emergency approval to COVID-19 vaccines for greater profits. Close to the US elections, the political war-of-words also kicked in, where anti-Trump users said they were not ready to trust a vaccine announced by Trump, calling it rushed, and pro-Trump users called them anti-vax. Opposing political blocs were also seen pointing fingers at each other's measures to fast-track the vaccine.   %Interestingly, Pfizer announcement came within a week after US elections 2020. Opposing geopolitical camps were also seen pointing fingers at each other's measures to fast-track the vaccine, possibly enforcing the idea of vaccine nationalism and eroding the trust further over the process (please see Supplementary Information for anti-vax discussion about China and Russia).
The major concerns discussed under `rushed vaccine' are about an experimental vaccine which was made in less than a year and the long-term side effects are not known. Typical process of multiphased (at least 3 phases pre-registration) clinical trials for new vaccine may take 10 years or more\footnote{\url{https://www.cdc.gov/vaccines/basics/test-approve.html}}. However, there was an obvious need to fast-track the process for COVID-19 vaccine. Sputnik vaccine was announced on Aug.\ 11, 2020, after phase 1 and 2 of the trials and phase 3 trials commenced post-registration \cite{Sputnik} and reported over 90\% efficacy. Pfizer vaccine was announced on Nov.\ 9, 2020 after the first interim data analysis of phase 3 trials, and they got FDA's EUA (Emergency Use Authorisation) on Dec.\ 11, 2020 after conclusion of phase 3 trials. 

Jokes (side effects) are discussed later under \hyperref[memes-sec]{Memes} because they appear to have a somewhat different dynamic.

%\vspace{-3mm}
\subsubsection*{Vaccine Mandate}

The single most significant topic discussed in anti-vax tweets is `vaccine mandate'. %As shown in Fig.~\ref{topics_figure}, this topic emerged as a central topic early in the yearlong observation period, saw a steady growth in the volume of tweets, and contributed to some of the biggest peaks in anti-vax tweet activity as discussed above. %Here, we discuss the various issues raised within the `vaccine mandate' topic. 
The expressed opinions ranged from mandates infringing on basic human freedom (if vaccination was required to travel, shop, dine, be employed, etc.) to the use of mandates for control and profit. For instance, some tweets suggested that coronavirus was purposely released, or the media was exaggerating the pandemic, to create business opportunities for the pharmaceutical companies or to limit people's rights as a secret plan to create an Orwellian totalitarian dystopia. Moreover, the virus control measures, such as, masks and lockdowns, are described as instruments for establishing control and coercing people into accepting vaccine. 

On Aug.\ 21, 2020, WHO's director said in a media briefing: "we cannot go back to the way things were" \cite{WHO-Aug21-2020}. Several tweets referred to this statement to suggest that COVID-19 is about changing society, and it is not about a virus. These tweets contributed to the vaccine mandate peak on Sep.\ 3, 2020. Protests and vaccine refusals from America's frontline doctors\footnote{\url{https://aflds.org/}}, a group of physicians against vaccine, and other medical professionals are also discussed in many tweets \cite{LATimes-refusals}. Later, we try to separate conspiracies and falsehoods from the discussion around genuine concerns to understand the composition of the anti-vax discourse.

We also found discussions about historical debates around parental consent within the context of childhood vaccinations, the slogan of `my body my choice' typically associated with abortion rights movements\footnote{\url{https://www.amnesty.org/en/get-involved/my-body-my-rights/}}, efficacy issues of influenza vaccines, and the comparison between voter's ID and vaccine certificate. Some tweets argue that taking a fast-tracked vaccine without fully understood long-term effects can only be an individual's choice. Other tweets suggested that making vaccines mandatory is not a solution, and we should just learn to live with the virus. 

%\vspace{-3mm}
\subsubsection*{Vaccine Efficacy (No Need for Vaccine, Ineffective Vaccine)}

We grouped `no need for vaccine' and `ineffective vaccine' into a theme of vaccine efficacy. Surprisingly, this theme did not get major traction during our study period, although, vaccine efficacy may have become a more significant talking point in the second year of the pandemic, with the emergence of the Delta and Omicron variants of COVID-19. 

The tweets on `ineffective vaccine' mainly highlighted and propagated the doubts raised by authorities, who at the time were uncertain if vaccination alone could resolve the pandemic. On Dec.\ 28, 2020, a WHO expert said: "At the moment, I don’t believe we have the evidence on any of the vaccines to be confident that it’s going to prevent people from actually getting the infection" \cite{WHO-Dec28-2020}. Many tweets simplified this statement to "there is no evidence that vaccine will prevent infection". These tweets have basically exploited the uncertainties associated with the vaccine efficacy, and have taken the statements out of context in line with the anti-vax narrative. The only significant peak in `ineffective vaccine' occurred on Jan.\ 15, 2021 similar to the tallest peak in the `side effects' topic, and is due to a meme about a zombie apocalypse resulting from a failed vaccine in the movie `I am Legend'. 

On `no need for vaccine', some tweets focused on the advice about using masks after vaccination, inferring (wrongly) that vaccines are pointless if masks are still needed. Moreover, the discussions in this topic mostly compared COVID-19 to influenza and the common cold, and claimed COVID-19 had a greater than $99.7\%$ survival rate. Multiple peaks in `no need for vaccine' topic occurred during Dec.\ 7-20, 2020 when the vaccines from Pfizer and Moderna were approved by the FDA~\cite{Pfizer-eua,Moderna-eua} (see discussions about Dec.\ 7-20, 2020 above).

%\vspace{-2mm}
\subsection*{The Common Topics}
\label{sec:common}

Pro- and anti-vax tweets discussed the common topics from opposite viewpoints as illustrated below.

%\vspace{-3mm}
\subsubsection*{Masks, Lockdowns, etc.}

%Public views about masks, lockdowns, and social distancing are probably as polarised as they are about COVID-19 vaccine, but that would be outside the scope of this paper. 
The topics of `masks' and `lockdowns, etc.' (which includes social distancing) are discussed in the context of vaccines in our dataset. Frequently raised issues in anti-vax tweets included (1) why a vaccine is needed if masks, lockdowns, and social distancing are effective, (2) why masks are necessary if vaccines are effective, (3) masks and lockdowns violate basic human rights, and are used to control the population and create need for vaccination, (4) tweets discussing how leaders and figureheads who were against masks are now in priority queues for vaccines. 

In contrast, the pro-vax tweets were overwhelmingly supportive of masks, lockdowns, and social distancing, discussing their importance in controlling the spread of virus and saving lives in the pre-vaccine period. %Considering that masks and lockdowns were the major measures to mitigate and control the spread of COVID-19 during our observation period, it is unsurprising that a significant number of tweets mentioned these measures while discussing COVID-19 vaccines. 
For anti-vax tweets, the most significant peak in `masks/lockdowns' in Fig.~\ref{anti-topics} came right after the Sputnik announcement (see above). The only significant peak in the pro-vax `masks/lockdowns' topic in Fig.~\ref{pro-topics} is due to the following viral tweet on Jun.\ 2, 2020,
%\vspace{-2mm}
\begin{displayquote}
{\it "Curfew for black lives matter but none for coronavirus..... black people can’t be more dangerous than an airborne virus with no vaccine."}
\end{displayquote}

%\vspace{-4mm}
\subsubsection*{Pfizer/Moderna}

The majority of the activities on this topic is in the post-vaccine period, and many peaks in Fig.~\ref{topics_figure} coincide well with relevant current events, including the Sputnik vaccine announcement and approval of the Pfizer and Moderna vaccines in the UK and USA.

%\vspace{-3mm}
\subsubsection*{Big Pharma}

Pandemic profiteering by pharmaceutical companies is a major issue discussed in anti-vax tweets under `big pharma'. They include significant increases in profit for Pfizer (and its partner company BioNTech) and Moderna~\cite{Pfizer-annualreport2021,Moderna-profits}, instead of supporting open licences to boost vaccination delivery in countries with low uptake rates. The famous reply by Polio vaccine inventor Jonas Salk, ``could you patent the sun?'' in response to why he did not patent his invention, was quoted in over two thousand tweets. Profiteering is generally frowned upon in both anti-vax and pro-vax tweets, although pro-vax tweets are mostly positive about the role of pharmaceutical companies in developing COVID-19 vaccines. Some anti-vax tweets, however, suggested the pandemic was purposely created for business and power grabbing opportunities. For example, the `big pharma' anti-vax tweets peaked on Sep.\ 24, 2020, due to the news about the UK's chief scientific adviser's \textsterling600,000 worth of shares in GlaxoSmithKline~\cite{Vallance}. 

%\vspace{-2.5mm}
\subsubsection*{Bill Gates}\label{gates}

The discussion about Bill Gates is prominent in the pre-vaccine period, with almost 96\% of the anti-vax tweets on Apr.\ 9, 2020 being on this topic (see `Bill Gates' in Fig.~\ref{anti-topics}). This activity caused a noticeable spike in total anti-vax tweets (see Fig.~\ref{result_Lamsal}), and a likely driver is the appearance of Bill Gates on Trevor Noah's The Daily Show on Apr.\ 2, 2020. The second peak is due to a surge in tweets on Aug.\ 8, 2020 after Bill and Melinda Gates funded \$150 million to the Serum Institute of India and the GAVI vaccine alliance for vaccine development~\cite{Serum-funding}. News, which appeared in tabloid press, about Elon Musk calling Bill Gates a `knucklehead'\footnote{\url{https://nypost.com/2020/09/29/elon-musk-says-he-wont-take-coronavirus-vaccine/}} was shared many times and caused a peak on Sep.\ 29, 2020.% (see Fig.~\ref{anti-topics}).

Besides these peaks, many anti-vax tweets covered various conspiracy theories, such as the pandemic being orchestrated to create a market for vaccines, often linked to Gates' 2015 TED Talk on the next epidemic outbreak or the global public health efforts supported by the Bill and Melinda Gates Foundation% to control and eradicate diseases
. Other conspiracies, such as `microchip in a vaccine' and `eugenics', were also directed to Bill Gates and his foundation's support and interests in e-vaccine cards. % and public health. 
His subsequent interviews and statements to clarify his role in vaccine development appeared only to amplify anti-vax tweet rhetoric. 

There was little interest from the pro-vax side in this topic (1.1\% or 399,560 pro-vax tweets). Pro-vax tweets, largely, shared news and updates about Bill and Melinda Gates funding and efforts in vaccine development. Some tweets also ridiculed the conspiracies associated with Bill Gates, especially, about microchip in the vaccine.

%\vspace{-3mm}
\subsection*{The Pro-vax Tweets}\label{sec:pro}

Of the top 10 pro-vax topics shown in Fig.~\ref{pro-topics}, `{\bf vaccine development}' is dominant throughout the study period. Over 50\% discussed this topic, including updates from clinical trials, news about the efforts of governments to fund the research and acquire vaccines for their people, vaccine rollout details, scientific information about the vaccines - especially mRNA technology, news about public leaders and celebrities making donations and publicly receiving a vaccine, etc. The topic plot shows various peaks which can be attributed to a combination of different causes, such as press releases about clinical trials - including trial results and issues, important vaccine announcements, updates about approval procedures, and the launch of various contact tracing apps. We observe peaks in `vaccine development' on Aug.\ 12, 2020 and during Dec.\ 7-20, 2020 (for the particular vaccine-related events, see \hyperref[sec:anti]{The Anti-vax Tweets}). One peak is attributed to the start of the Indian vaccination program on Jan.\ 16, 2021. The US Presidential Debates were also referenced in the context of `vaccine development', especially in the weeks leading up to the elections. 

The topic `{\bf debunking anti-vax conspiracies}' also shows multiple peaks during Dec.\ 14-20, 2020, which is due to a viral meme mocking anti-vax concerns of the ingredients in vaccines. This topic mostly contains humorous tweets making fun of anti-vax conspiracies, such as, vaccine containing microchip or government trying to kill people using vaccine, etc. 

The topic of `{\bf pro-vax (general)}' is the second most popular pro-vax topic with 9\% tweets and include positive statements about COVID-19 vaccine and vaccine in general, such as, people posting after getting COVID-19 vaccine, sharing updates about vaccine rollout and eligibility, encouraging other people to get the vaccine, etc. There is significant overlap between the topics of `vaccine development', `distribution', `{\bf setbacks and issues}', `{\bf access to vaccine}', `pro-vax (general)', and `debunking anti-vax conspiracies'. Altogether, they define almost $70\%$ of the pro-vax tweets. % covering the development and distribution of COVID-19 vaccine, trial and rollout issues, and in general encouraging and supportive arguments for vaccine uptake. 
The rest of the pro-vax topics are discussed in the Supplementary Information.

%\vspace{-3mm}
\subsection*{Memes} \label{memes-sec}

\begin{table*}[tbh!]{
  \begin{center}
    \caption{Memes among high frequency anti-vax and pro-vax tweets}%\vspace{-2mm}
    \label{memes_table}\vspace{-3mm}
    %\resizebox{!}{1.2cm}{
    \begin{tabular}{l l c c c}
	\hline \hline
	 & {\bf Original tweet} and example variations & Start date & Variants& Tweets\\
	\hline \hline
	\parbox[t]{2mm}{\multirow{18}{*}{\rotatebox[origin=c]{90}{Anti-vax (from top 10 tweets)}}} & {\bf Me after taking the Covid-19 vaccine *funny picture*} & Nov 9, 2020  & 103 & 245658 \\
	&  Me and the girls once we get the vaccine *funny picture*, & & &  (2.3\%)\\
	& Me and the boys when we get the Covid vaccination *funny picture*,& & & \\
	& snoop dogg after the vaccine *funny picture*, etc. & & & \\
	\cline{2-5}
	& {\bf when that vaccine hits *funny picture*} & Dec. 8, 2020 & 34& 71438\\
	& the gang and i after that vaccine hits *funny picture*, & & & (0.7\%)\\
	& Me and the girls when the vaccine hits *funny picture*, etc. & & &\\
	\cline{2-5}
	& {\bf he is not lying, this is me before and after the vaccine. not enough people} & Jan. 17, 2021 & 1& 38492\\ 
	& {\bf are speaking up about this *funny picture*} & & &  (0.4\%)\\
	\cline{2-5}
	& {\bf I Am Legend was set in 2021... The zombie apocalypse was because of the} & Jan. 14, 2021& 1& 38355\\
	& {\bf  failed vaccine.....}& & & (0.4\%)\\
	\cline{2-5}
	& {\bf i just saw 3 homeless guys giving each other the covid vaccine under a} & Jan. 13, 2021  & 2 & 38331 \\ 
	&   {\bf bridge. what a caring community we live in} & &   & (0.4\%)\\
	& "just saw 4 homeless men giving eachother the covid vaccine under a bridge",  & & & \\
	& "what a caring community we live in"& & &\\
	\cline{2-5}
	& {\bf why they warming my vaccine in a spoon} & Jan. 18, 2021 & 3& 37510 \\
	& how come they heated my vaccine up in a spoon?, etc.& & &(0.4\%)\\
	\cline{2-5}
	& {\bf this is the lab that invented the coronavirus vaccine *funny picture*} & Nov. 11, 2020 & 17& 35857\\
	& This is the vaccine *funny picture*, & & & (0.3\%)\\
	&  This is what is wrong with the AZ vaccine, etc. & & &\\
	\hline 
	\parbox[t]{2mm}{\multirow{12}{*}{\rotatebox[origin=c]{90}{Pro-vax (from top 15 tweets)}}} 	& {\bf You eat sausages your whole life but you refuse vaccine because you don't}  &  Nov. 20, 2020 & 254& 245051\\
	& {\bf know what’s in it.} & & & (0.7\%) \\
	& if u eat jack in the box tacos don't  worry about what's in the covid vaccine, & & &   \\
	& If you ate these as a kid. You don't have to worry about the vaccine, etc. & & &\\
	\cline{2-5}
	& {\bf n****s is more scared of the vaccine than the virus} & Dec. 5, 2020 & 4 & 75222\\
	& How y'all scared of the vaccine but not scared of Covid, & & & (0.2\%)\\ 
	%& All the drugs y’all do but y’all scared to take a vaccine, etc. & & &\\
	\cline{2-5}
	& {\bf My boyfriend got his covid vaccine yesterday and I can tell you the most} & Dec. 16, 2020& 1& 72117\\
    & {\bf   prominent side effect is the inability to shut up about getting the covid vac}& & & (0.2\%)\\
	\cline{2-5}
	& {\bf we gotta get this goddamn vaccine} & Nov. 25, 2020 & 2& 45752\\
	&  when we get the vaccine & & & (0.1\%)\\
    \cline{2-5}
	& {\bf Y'all swear the government need a vaccine to kill y'all like Burger King}  & Dec. 19, 2020 & 1& 37336 \\
	&{\bf don't sell 10 nuggets for \$1.50} & & & (0.1\%)\\
	\hline \hline
	\multicolumn{5}{l}{Some of the most frequent retweeted tweets in both anti-vax and pro-vax groups are memes or funny tweet, specially in the}\\
	\multicolumn{5}{l}{post-vaccine era. This table lists the memes along with the related variants found among the top 10 anti-vax and the top 15}\\
	\multicolumn{5}{l}{pro-vax tweets. What we see here is a very interactive tête-à-tête between almost similar strength anti-vax and pro-vax groups.}\\
	\multicolumn{5}{l}{In relative terms, memes are more significant for anti-vax group. Please see Supplementary Information for examples of funny}\\
	\multicolumn{5}{l}{pictures referred here.}
	\end{tabular}
 \end{center}}
 %\vspace{-9mm}
\end{table*}

Memes or humorous tweets, with occasional variations, are among the most frequently retweeted in both anti- and pro-vax groups. our results are summarised in Table~\ref{memes_table}. These funny messages, although not carrying any argument or information, are perhaps more memorable and shareable. Of the top 10 most retweeted anti-vax tweets, 7 are jokes or memes with multiple variants, which constitute almost $5\%$ of the total volume of anti-vax tweets. We also examine the top 15 most retweeted pro-vax tweets (the top 10 have only 2 memes) and find 5 different jokes and memes. Altogether, they account for $1.3\%$ of the total pro-vax tweets. 

Determining whether memes and humour were deliberate attempts to advance the anti-vaccine discourse, %or just more prominent as such messages are more shareable on social media, 
is an important question but one beyond the scope of this paper. What is perhaps remarkable is that, based on the timeline identified in Table~\ref{memes_table}, there almost appears to be a reactionary tête-à-tête between anti-vax and pro-vax memes, triggered by a sequence of news events around vaccine approvals and development (see the events 3-5 % identified for Dec.\ 7-20, 2020, Dec.\ 28, 2020-Jan.\ 3, 2021, and Jan.\ 15-17, 2021 
in \hyperref[sec:anti]{The Anti-vax Tweets}). 

%\vspace{-3mm}
\subsection*{Genuine Concerns vs.\ falsehoods in Anti-vax Discourse}\label{sec:legit}

\begin{figure}
\begin{center}
\includegraphics[width= 0.4\textwidth,height=4.2cm, clip = true]{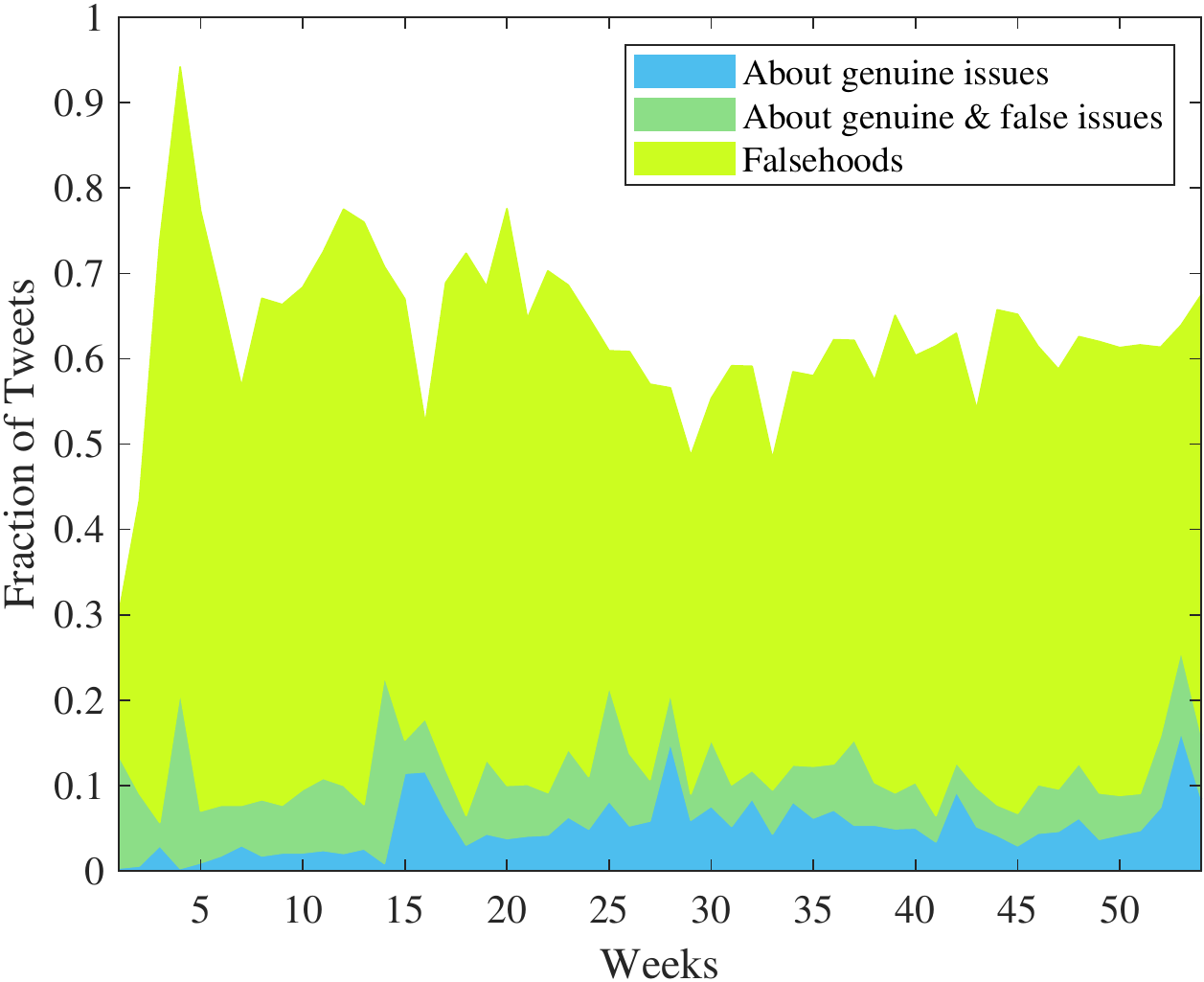}\vspace{-2mm}
\caption{Besides the conspiracies about mRNA vaccine altering DNA, etc., a lot of misinformation and falsehoods found its way into all anti-vax topics, such as, side effects, vaccine mandate, etc. We filter the tweets which discussed some genuine issue, though it may still contain misinformation. The `about genuine issues' plot shows that only 10-15\% of the anti-vax tweets referred to some legitimate concern. The 'falsehood' plot tracks the tweets about fictitious or spurious issues. The 'About genuine and false issues' plot shows the fraction of tweets which have keywords related to both genuine and false issues.}
\label{legit}
\vspace{-9mm}
\end{center}
\end{figure}

A key problem of interest is to understand the contribution of misinformation and falsehoods toward the anti-vax discussion. To investigate, we separate the keywords/key-phrases, used for anti-vax topic classification, into three groups: (1) genuine issues, (2) falsehood, and (3) neither. Then, we count the number of anti-vax tweets with topics appearing in the first two groups, as the `neither' group could also be put together with the anti-vax tweets not classified into any topic. It is possible that one tweet can be classified into multiple groups, as shown by the plot of `About genuine and false issues' in Fig.~\ref{legit}, as the tweet may contain keywords related to genuine issues and falsehoods.

The topics of genuine or legitimate concerns are identified by consensus among the authors, as detailed in the Supplementary Information. We consider these topics genuine concerns as they offer some grounds for reasoned debate and argumentation, as opposed to demonstrably false claims or fictitious issues where no reasoned discussion is possible. The genuine concerns are (see Supplementary Information for details)
\vspace{-2mm}
\begin{itemize}
    \setlength\itemsep{-0.5em}
    \item Mandatory vaccines and loss of freedoms
    \item Fast-tracked vaccine
    \item Historical issues with vaccines and clinical trials
    \item Pharmaceutical companies profiteering
    \item General vaccine side effects
    \item Blood clots after receiving the AstraZeneca vaccine
    \item Waning immunity and virus variants
    \item Administrative mismanagement
    \item Animal abuse during vaccine development.
\end{itemize}\vspace{-2mm}

Conversely, topics found under falsehoods include conspiracies, anti-vax memes, and tweets about fictitious issues. The conspiracies we identify were: `experimental mRNA/altering DNA', `vaccine contains microchip', `metal/nanoparticles in vaccine', `cells from aborted babies in vaccine', `fertility issues', `Robert F. Kennedy' (as a figurehead in the vaccine scepticism movement), `related to flu vaccine', `eugenics', `vax makes you gay', and `QAnon'. The Supplementary Information provides a detailed breakdown on this group of 10 `conspiracies'. Fictitious issues include blaming COVID-19 vaccines for the deaths of 23 elderly people in Norway, claims that COVID-19 fatality rates were lower in comparison to influenza, claims that COVID-19 has naturally diminished without a vaccine, or that there was no point to have vaccines. The tweets which are not classified into either `genuine issues' or `falsehood' are typically simple statements such as ``I do not want a vaccine'', or accusations at public leaders because of something they said or did, discussions about alternative treatments being a better choice than a vaccine, and also some tweets which are not classified because of lack of suitable keywords.  

Figure \ref{legit} shows the fraction of tweets under each group for each week throughout the year. Only 10-15\% of anti-vax tweets contained some genuine issue, while a large portion was based on fictitious issues and falsehoods (although `conspiracies' accounted for just $6.6\%$ of the anti-vax tweets). Around 38\% of the yearlong anti-vax tweets belonged to neither of the groups. We remark that our keyword-based classification cannot distinguish between well- and poorly-reasoned tweets about a genuine topic. Many tweets identified as discussing a `genuine concern' in fact used excessive exaggeration or presented information about true facts/events out-of-context, such as the following tweet.% that referred to a previous vaccine incident. 
%\vspace{-2mm}
\begin{displayquote}
{\it "The last rushed vaccine cost the government £60M in payments for compensation of injury. How long will it be before we hear about the side effects of a C19 vaccine when its rolled out to millions? Humanity has lost it's way in thinking a poisonous concoction is `for our safety'."}
\end{displayquote}\vspace{-2mm}
Nonetheless, there are some well-reasoned tweets in this group as well, such as \vspace{-2mm}
\begin{displayquote}
{\it "Take the jab or lose your job: Medical journal calls for a MANDATORY Covid vaccine, says `noncompliance should incur a penalty'."}
\end{displayquote}\vspace{-2mm}

From Fig.\ \ref{legit}, falsehoods clearly dominate the topics discussed in anti-vax tweets, although there is a slight decline towards the end of the observation period. While our findings around dual-stance users suggests that the users on Twitter are not as polarised around COVID-19 vaccines as one may initially assume, the significant amount of falsehoods found in anti-vax tweets does confirm that addressing online misinformation around vaccines is still a critical challenge.

%\vspace{-3mm}
\section*{Discussion}

In this paper, we used NLP-based stance detection and composite topic modelling to examine a Twitter dataset containing $75$ million English tweets relating to COVID-19 vaccines, classifying tweets into `pro-vax' and `anti-vax' stances and exploring leading factors behind each stance. We examined the unique user IDs in each group and found a power law distribution describing the number of users with a given number of tweets, indicating a significant contribution to the discussion from a small number of users. Moreover, we found close to $2$ million dual-stance users, who had tweeted both anti-vax and pro-vax tweets. Our topic analysis revealed the core topics discussed by the `anti-vax' and `pro-vax' tweets, including a set of common topics. Anti-vax discussion was spread across multiple topics, such as, vaccine safety and mandates, etc. Falsehoods featured significantly more than discussion in reference to genuine concerns in almost all of the anti-vax topics. On the other hand, pro-vax discussions centred primarily around vaccine development. Peak tweet activity often coincided with a key COVID-19 current event or media report, while memes and jokes featured heavily in the most popular retweeted messages.

Our methods and findings fill gaps in and extend the literature about COVID-19 vaccine discussions in online social media. NLP and transformer-based stance detection with a training set of $42,000+$ labelled tweets should provide a more accurate reflection of tweet stance, and hence the overall nature of COVID-19 discussion on Twitter, compared to existing literature that use sentiment analysis and other unsupervised learning methods. Our approach complements extant works that have explored the spread of (mis)information over Twitter and other social media~\cite{Cinelli2020}, by examining the key topics discussed and differentiated by stance. We are also able to track the topic evolution over an entire year, again differentiated by stance, identifying topics that have a sustained presence in the Twitter discourse for both anti-vax and pro-vax tweets (e.g. `vaccine mandate') and those topics that emerge and disappear (e.g. `vax makes you gay', see Supplementary Information) --- this contrasts existing works which focus on specific topics or narrow timelines at perhaps a higher resolution~\cite{Lanyi2022,Nazar2021}. Trust and safety-related topics, e.g. `side effects', `rushed vaccine', featured prominently in the anti-vax tweets, which supports other work that identified safety and trust (in institutions and governments) as a key hurdle in addressing vaccine hesitancy~\cite{Lanyi2022,Kucukali2022,Hou2021}. Our stance detection approach also enabled a definitive identification of a set of `dual-stance' users who contributed a significant volume to both anti-vax and pro-vax tweets, supporting existing findings~\cite{Gori2021}.

Nonetheless, our study has certain limitations, which we hope to address through future work. First, the Lamsal dataset, from which we extracted the data we analysed, contains only English tweets and hence our findings are most relevant to nations with English as the predominant language (e.g. USA, the UK, Australia, Canada) --- we did not attempt to differentiate based on location as Twitter geolocation data can be unreliable. It would be of interest to conduct similar stance detection-based analyses for tweets in other languages, perhaps utilising cross-lingual transfer learning techniques \cite{Conneau2019}. Second, we have only considered Twitter, and it would be of interest to apply our trained model on datasets from other social media platforms such as Reddit. Third, we were not able to hydrate all tweet IDs from the Lamsal dataset (e.g. tweets by Donald Trump and their retweets are not visible after his Twitter account was banned). 

Our study lays the groundwork for several promising directions of future work. Obviously, it would be of interest to extend our analysis to cover the second year of the COVID-19 pandemic, as our observation period ended as COVID-19 vaccine roll-outs were accelerating in many countries. More in-depth analysis of the users, including especially dual-stance users who contribute significantly to the overall Twitter discussion environment, would complement the current focus on topic modelling. A closer examination of how anti-vax and pro-vax topics interact, including determining causality relationships, would also be of interest, e.g. identifying whether peak activity in a pro-vax topic led to peak activity in some anti-vax topic. 

%\vspace{-3mm}
%TC:ignore
\section*{Methods}\label{methods}
\subsection*{Dataset}

This study uses the publicly available dataset collected and maintained by R. Lamsal~\cite{Lamsal2021}, which collected the tweet IDs for global English language tweets that featured COVID-19 related keywords. The complete list of keywords can be found in Ref.~\cite{Lamsal2021}. We started hydrating Lamsal's dataset in Feb.\ 2021, which involved retrieving the tweet and associated attributes from the tweet ID in the Lamsal dataset. We observed that $20-40\%$ of tweets could not be retrieved for the period Mar.-Apr.\ 2020. This hydration loss gradually came down to $5-15\%$ towards Feb.-Mar.\ 2021. In order to extract the vaccine related tweets, we further filtered the dataset with the keywords `vaccine', `vaccination', `vax', `vac', `jab', and `shot'. 

%\vspace{-2mm}
\subsection*{Stance detection}

Our study used a stance detection tool created from OpenAI's GPT transformer model \cite{Rao2019} to classify tweets into three categories: favour, against, and none, for the topic `vaccine hesitancy'. The classification accuracy, measured in terms of composite F-score for favour and against classes~\cite{Rao2019}, depends critically on the labelled set used to fine-tune the GPT model. After various trials, we ended up selecting around 100 tweets per day, which gave us reasonable F-scores of 0.6 or above for 20 test sets we picked and labelled from the yearlong dataset~\cite{Lamsal2021}. Test sets are comprised of 250-1000 randomly selected tweets from day 1, 20, 40, 60, ..., from the 368 days in the dataset. The composite F-scores for the 20 test sets are in the range of 0.67-0.87. 

In order to make this yearlong study possible with reasonable confidence, we had around $42,000+$ tweets to label. As the conversation was changing throughout the year, it was critical to have a relevant labelled set of tweets to fine tune the GPT model, which can then be used for detecting stances for the tweets of that specific period. Considering the enormous task load, we devised an annotation strategy with two phases. A primary annotator provided one set of labels for all $42,000+$ tweets in the first phase. The second phase focused on estimating the bias error in annotation by the primary annotator from the first phase. Extra annotators provided at least two additional sets of independent labels for 3000 tweets from Mar-Apr 2020, and 3000 tweets from Sep-Oct 2020, and a majority rule was applied to get the final label. There was a 93\% similarity between the final labels and primary annotations for Mar-Apr 2020 and 89\% for Sep-Oct 2020. We remark that the period Sep-Oct 2020 was exceptionally complex for annotation due to an ongoing contentious debate between Democrat and Republican voters in the US, and leaders from both sides were called anti-vax. We labelled all such political tweets with the stance `none', noting that it was not always clear if the main focus of the tweet was vaccines or politics.

The stance prediction results when the GPT model was fine-tuned using both sets of labels, i.e., final labels and primary annotation only, are 91\% similar for the Mar-Apr 2020 period and 84.1\% for the Sep-Oct 2020 period. Both set of labels result in almost similar predictions. We thus argue that the bias error is marginal in this study and can be safely ignored. Moreover, any misclassification of tweets due to personal bias can be handled, along with the prediction error of the GPT model, when each class is thoroughly analysed for specific discussion points through the subsequent topic modelling.

%\vspace{-2mm}
\subsection*{Topic modeling} 

LDA topic modelling relies on co-occurrence of words in a document and a major issue with tweets, which comes under Short Text Topic Modelling (STTM), is that it is limited to a maximum of 280 characters~\cite{Qiang2019}. We tested several topic modelling tools, such as, LDA\footnote{\url{https://nicharuc.github.io/topic_modeling/}}, GS-DMM \cite{Yin2014}, Bertopic \cite{Grootndorst2022}, Top2vec \cite{Angelov2020} --- we found that GS-DMM, an LDA based topic modelling tool optimised for STTM, was most suitable in finding relatively distinct and meaningful cluster of words. However, GS-DMM alone was not sufficient for the topic classification tasks, because as it turns, the core topics are highly intertwined with one another and use similar words. 

We, then, decided to use GS-DMM for keyword collection instead and added another phase in our strategy where these keywords are manually searched in the tweet files and key-phrases are selected for final topic classification. The key-phrases lowered the misclassification rate considerably. We manually checked the 50 high frequency tweets under each topic to calculate the error rate. More than half of the topics have zero error, and only 7 topics have an error rate higher than $10\%$. The topic of `RNA altering DNA' has the highest error rate of $25\%$, `side effects' has $19\%$ error, and $16.8\%$ error was seen in the topic of `microchip' but these errors appear to be mostly due to stance misclassification of satirical tweets. The topic `Bill Gates' has an error rate of $14.7\%$ for the anti-vax class, and this is due to the stance detection tool misclassifying tweets that discussed legislative bill and Bill Clinton.

%\vspace{-3mm}
\subsection*{Network Graph}

The two main tools used to build a retweet network are NetworkX and Gephi.
In a retweet network, users are represented by nodes, and retweet relationships are represented by edges. Edges are directed. The retweeted users and the original tweeted users in the database are extracted to construct an edge table. Then NetworkX is used to convert the edge table into a network file. Gephi was used to visualise the network file. 

%TC:endignore
%\vspace{-3mm}
\subsection*{Data Availability}{The code used and the data generated in the simulations is available at \url{https://github.com/https://github.com/ZainabRZaidi/COVID-19-Vaccine-Hesitancy-}}.

% Bibliography
%\bibliographystyle{pnas}
\bibliography{reference}

\section*{Acknowledgement}

We are thankful to Adeel Razi, Fatemah A. Husain, Rabindra Lamsal, and Lavy Libman for their discussions and advice regarding different aspects of the study.
\vspace{-2mm}

\section*{Author contributions statement}

ZZ, MY, SK, YK designed research; ZZ, FJS, AJ, BG performed research; ZZ, MY, FJS, AJ, BG, CG, JE, SK, YK analyzed data; ZZ, MY, YK wrote the paper with inputs from SK.

\end{document}